**Title: Crystallization induced Sulfur and REE zoning in apatite: the example of the Colli Albani's magmatic system**

Running title: Zoning in apatite from the Colli Albani

Alessandro Fabbrizio[a] (Corresponding Author)

[a]Institute of Petrology and Structural Geology, Faculty of Science, Charles University, Albertov 6, 12843 Prague, Czech Republic

Email: alessandro.fabbrizio@natur.cuni.cz Tel: +420 22195 1525

Mario Gaeta[b]

[b]Dipartimento di Scienze della Terra, Sapienza, Università di Roma, Italy

Email: mario.gaeta@uniroma1.it

Michael R. Carroll[c]

[c]School of Science and Technology, Geology Division, Università di Camerino, 62032 Camerino, Italy

Email: michael.carroll@unicam.it

Maurizio Petrelli[d]

[d]Department of Physics and Geology, University of Perugia, Piazza dell'Università, 06123 Perugia, Italy

Email: maurizio.petrelli@unipg.it

**Abstract**

We investigate the distribution of major and trace elements in apatite crystals hosted in granular alkaline rocks composed mainly of leucite and clinopyroxene, representative of the hypabyssal crystallization of a magma body in the Colli Albani Volcanic District (CAVD). The Colli Albani is a Quaternary ultra-potassic district that was emplaced into thick limestone units along the Tyrrhenian margin of Italy. Results show that the analyzed crystals are the SrO-richest (up to 4.6 wt.%) fluorapatite (F = 2.6-3.7 wt.%) of the Italian alkaline rocks. The strontium enrichment is caused by the lack of other Sr-compatible mineral phases, such as plagioclase, alkali feldspar, and melilite in the mineralogical assemblage of these leucite- and clinopyroxene-bearing rocks and by Sr assimilation from the carbonate wall rock. The studied samples show core-rim zoning with rims enriched in Si, S, and REE whereas the cores are enriched in Ca and P. The LREE-oxides contents of apatite, reaching up to 4.2 wt.%, represent more than 95% of total REE budget, $SiO_2$ contents range from 1.3 to 3.6 wt.%, and $SO_3$ concentrations are between 0.6 and 1.4 wt.%. We support the idea that the REE and sulfur enrichments are a response to the crystallization caused by the pressure drop in the magmatic system during the eruption and follow the substitution mechanisms $Si^{4+} + REE^{3+} = P^{5+} + Ca^{2+}$ and $S^{6+} + Si^{4+} = 2P^{5+}$. These results also suggest the presence of the substitution $Th^{4+} + 2Si^{4+} = Ca^{2+} + 2P^{5+}$.

Keywords: Apatite; Colli Albani; zoning; REE; sulfur; crystallization

**Introduction**

Apatite, due to the openness of its crystal structure, can incorporate significant amounts of rare earth elements (REE) (Ito, 1968; Larsen, 1979; Watson & Green, 1981; Roeder *et al.,* 1987; Rønsbo, 1989; Fleet & Pan, 1995; Pan & Fleet, 2002). As a consequence, it plays an important role in controlling REE variations in igneous systems. Apatite also provides a useful petrological tool that can give crucial information on the crystallization history of igneous complexes (Rønsbo, 2008). As an example, natural apatite, such as those from the Ilímaussaq alkaline complex, can be enriched to more than 20 wt.% $REE_2O_3$ and the REE enrichment can be balanced by at least two different substitution mechanisms: $Ca^{2+} + P^{5+} = REE^{3+} + Si^{4+}$ and $2Ca^{2+} = Na^{+} + REE^{3+}$ (Rønsbo, 1989, 2008).

In addition, apatite can incorporate appreciable amount of sulfur into the tetrahedral site as the $S^{6+}$ ion following three different substitutions schemes: $S^{6+} + 2Si^{4+} = 2P^{5+} + 2Ca^{2+}$, $S^{6+} + Na^{+} = P^{5+} + Ca^{2+}$, and $2S^{6+} + 4REE^{3+} + Si^{4+} + 2Na^{+} = 4P^{5+} + 5Ca^{2+}$ (Rouse & Dunn, 1982; Liu & Comodi, 1993; Parat *et al.,* 2002). S-rich apatite characterized by $SO_3$ contents up to 1.3 wt.%, with complex core-rim zoning have been found, for example, in the andesitic lava of the Eagle Mountain, USA, where $SO_3$ contents in apatite reflect the sulfur evolution in the magmatic system (Parat *et al.,* 2002).

Experimental studies, under oxidizing conditions, have been performed in order to use sulfur concentration in apatite as a geochemical tracer (Parat & Holtz, 2004, 2005). These studies were about the quantification of sulfur solubility in apatite and its distribution among S-rich apatite, melt, and the fluid phase (Parat & Holtz, 2004, 2005).

Regarding natural samples, normal and reverse zoning of sulfur in apatite microphenocrysts from the 2001 andesitic lava of Shiveluch Volcano, Kamchatka, have been correlated with $Na_2O$ contents, suggesting the presence of the two substitutions schemes involving the $Na^{+}$ ion. The observed sulfur zoning has been explained by differences in dissolved sulfur contents between intruding magmas (Humphreys *et al.,* 2006).

Recently, apatite has been used to determine the volatile contents of terrestrial and extraterrestrial magmatic systems (Piccoli & Candela, 1994; Boyce & Hervig, 2008; McCubbin *et al.,* 2011; Boyce *et al.,* 2014; Scott *et al.,* 2015; Stock *et al.,* 2016). Despite its great potential as a petrological tool, there were only few studies focused on apatite zoning. As an example, luminescence zoning that correlated U and REE concentrations in hydrothermal apatite crystals from a tin-tungsten deposit was reported by Knutson *et al.* (1985). On the basis of zoning profiles and the lack of chemical correlation among apatite grains sampled on a meter scale, the zoning was attributed to heterogeneous fluid compositions. Patchy REE zoning and variable zoning among adjacent apatite crystals in a pegmatite have been described and attributed to disequilibrium crystallization in a heterogeneous melt/fluid system (Jolliff *et al.,* 1989). Sector zoning in apatite grains from hydrothermal veins and pegmatites have been reported by Rakovan & Reeder (1994). Complex chemical zoning in apatite crystals from the Idaho batholith was attributed to changes in melt composition (Tepper & Kuehner, 1999). Zoning in several elements including Sr, REE, Th and U in apatite from the Shap granite, northern England, has been studied and related to the progressive crystallization within the magma chamber (Dempster *et al.,* 2003). Significant zoning in U and Th has been observed in Durango apatite used as standard for (U-Th)/He thermochronology, suggesting that the use of small crystals (hundreds of micron in length) as standards can significantly influence the thermochronology data (Boyce & Hodges, 2005). Oscillatory, sector and irregular zoning in apatite crystals from the Ilímaussaq alkaline complex revealed unusual details of events in successive stages of crystallization and mineral re-equilibration during the evolution of the magma chamber (Rønsbo, 2008). The effect of U-Th zonation from a large suite of apatite from the northwestern Canadian shield has been investigated, leading to the conclusion that it has negligible effects on the interpretation of (U-Th)/He thermochronometry data (Ault & Flowers, 2012).

Apatite with different degrees of enrichment in Si, REE, S, F, Th and U, reaching sometimes the britholite composition, is quite common in potassic and ultrapotassic volcanic districts of central and southern Italy (Orlandi *et al.,* 1989; Stoppa & Liu, 1995; Comodi *et al.,* 1999; Della Ventura *et al.,* 1999; Fulignati *et al.,* 2000; Oberti *et al.,* 2001; Melluso *et al.,* 2010, 2011). One of the best known is the Colli Albani Volcanic District (CAVD) developed along the Thyrrenian margin of Italy during Quaternary times. Its peculiar chemical, mineralogical and petrographical features have attracted the scientific community since the classical research by Washington (1906). Through the years, most scientific efforts in the Colli Albani have been focused on field relations, geochronological studies of eruptive events, mineral chemistry of rock forming minerals and whole-rock geochemical investigation of the individual rock types (*e.g.* Barbieri *et al.,* 1975; Peccerillo *et al.,* 1984; Ferrara *et al.,* 1985; De Rita *et al.,* 1995; Freda *et al.,* 1997; Gaeta, 1998; Gaeta *et al.,* 2000; Gaeta & Freda, 2001; Marra *et al.,* 2003; Boari *et al.,* 2009; Funiciello & Giordano, 2010; Di Rocco *et al.,* 2012; Gozzi *et al.,* 2014; Gaeta *et al.,* 2016). However, specific studies on the apatite mineral chemistry in order to understand its meaning in the magmatic evolution of the CAVD are relatively scarce (Federico & Peccerillo, 2002; Melluso *et al.,* 2010).

In this study, apatite crystals from leucite and clinopyroxene-bearing hypabyssal granular rocks of CAVD have been studied and analyzed by electron microprobe and LA-ICP-MS. The aim is to elucidate the processes involved in the creation of the considerable core-rim zoning for many elements (LREE Si, S, Ca, P, and F) and of the high SrO contents in the studied apatite crystals.

## Sampling & Analytical Methods

Apatite occurs in granular, leucocratic, lithic clasts characteristic of the main pyroclastic deposits of the Colli Albani volcanic district (Figs. 1, 2). In particular, analyzed rocks were sampled in the *Pozzolanelle;* the upper flow unit of the Villa Senni eruptive unit, emplaced at ≈365 ka, and related to the formation of Tuscolano-Artemisio caldera (Freda *et al.,* 1997; Marra *et al.,* 2009 and references therein). In agreement with the more recent literature regarding the CAVD, these granular, leucite and clinopyroxene-bearing, lithic clasts are called here italites (Trigila *et al.,* 1995; Dallai *et al.,* 2004; Gaeta *et al.,* 2006). This term even if coined for leucite-rich foidolites (Washington, 1920) describes better the absence of the K-feldspar and the plagioclase than other foidolite names (*i.e.* fergusites).

Analytical methods are fully described in appendix S1-S7 freely available online as Supplementary Material linked to this article on the GSW website of the journal, http://eurjmin.geoscienceworld.org/.

## Results

### *Petrography of Italites*

Tipically, italites belonging to the CAVD show a holocrystalline, allotriomorphic to idiomorphic textures, made up of millimeter-sized leucite and clinopyroxene, accompanied by submillimeter-sized phlogopite, nepheline, apatite and garnet (Fig. 3). The most abundant phases occurring in these rocks are clinopyroxene (Cpx) and leucite (Lct). Moreover, the italites show textures correlated with the modal abundance of these minerals. Optical microscopy observations show that allotriomorphic italites are characterized by similar content of leucite and clinopyroxene (Lct/Cpx≈1, *i.e.* CA10, CA25, CA28), while the idiomorphic italites, are characterized by the Lct/Cpx>1, *i.e.* CA24.

Variation diagrams, trace elements versus MgO, show that rock types of the CAVD are linked by a crystal fractionation process, where italite compositions represent the most primitive liquid (Freda et al., 1997). Moreover, indications of the $CO_2$ influx from the carbonate wall rocks are given by the occurrence in italites of garnet enriched in Ti, which is not stable in $CO_2$-free environments (Deer *et al.,* 1982), by the crystallization of phlogopite which indicates increasing $H_2O/CO_2$ ratio with differentiation in the magma reservoir, due to lower $CO_2$ solubility compared to $H_2O$ (Freda *et al.,* 1997), and by crystallization of leucite due to the relatively low $P(H_2O)$ (Freda *et al.,* 1997; Gaeta *et al.,* 2000). Furthermore, the composition of phlogopites confirm a crystallization environment that was influenced by the decarbonation of the host wall rocks (Gaeta *et al.,* 2000). Geochemical features represented by variation in oxygen isotope values associated with constant $^{87}Sr/^{86}Sr$ and LREE/HREE ratios, observed in clinopyroxene crystals belonging to different lithotypes (scoria clasts, glassy scoria clasts, cumulates and italites) from the Villa Senni Eruptive Unit are explained by assimilation of small amounts of sedimentary carbonate (Dallai *et al.,* 2004; Gaeta *et al.,* 2006). In addition, result of dating performed on leucite crystals in italite is in agreement with the age of the Villa Senni deposits, confirming that italites and Villa Senni products crystallized in the same magmatic system (Dallai *et al.,* 2004; Gaeta *et al.,* 2006). All these observations strongly suggest that italites represent the hypabyssal crystallization of the Colli Albani's magmas at the margins of the magma chamber (Freda *et al.,* 1997; Gaeta *et al.,* 2000; Dallai *et al.,* 2004; Gaeta *et al.,* 2006).

*Core-rim chemical zoning by EMPA*

Apatite crystals are euhedral with size ranging from 50 to 1000 μm in length (Fig. 3, Appendix S3). The analytical data (Appendix S2) indicate that we have reliable data (above detection limit) for P, Ca, Sr, Na, LREE, Si, S and F. BSE images and EMPA analyses indicate compositional zoning (core vs. rim) in all apatite crystals. The compositional differences between cores and rims are noteworthy. Total contents of LREE-oxides, vary from ~1 (core) to 4.2 wt.% (rim). The apatite crystal rims, compared to their cores, are enriched in La, Ce, Pr and Nd (Appendix S8). In all analyses, Ce is the dominant LREE with content up to 2.2 wt.%. Apatite are also characterized by significant amount of $SiO_2$ and $SO_3$ with a clear core-rim zoning (Appendix S8). Levels of $SiO_2$ increase from 1.3 (core) to 3.6 wt.% (rim) and those of $SO_3$ from 0.6 (core) to 1.4 wt.% (rim). Fluorine is the principal anion varying between 2.6 and 3.7 wt.%, in general the rims are enriched in fluorine respect to their cores (Appendix S9). A reverse zoning has been detected with a systematic reduction of CaO (from ~52 wt.% to ~49 wt.%) and $P_2O_5$ (from ~38 wt.% to ~34 wt.%) from core to rim (Appendix S9). Apatite are unusually enriched in SrO (see section 4.3), ranging from ~ 2.8 to ~ 4.6 wt.%, however no clear core-rim zoning has been revealed (Appendix S9).

*X-ray maps for S, La, Ce, and Nd*

Most of the apatite samples examined by X-ray intensity maps (Appendix S10; zoning also visible in BSE images, Appendix S3) display evident spatially related zoning patterns with cores poor in S, La, Ce and Nd surrounded by a narrow, generally S- and REE- rich. The presence of chemical zoning was also confirmed by the line profile analyses (Appendix S11).

The outer edges of the S- and REE- poor cores typically define anhedral shapes with evidence of resorption (crystals from samples CA24, CA25 and CA28) and in some case subeuhedral shapes (crystals from sample CA10) and are parallel to the outer edges of the crystal. Most of the crystals show a roughly uniform S- and REE-poor core. The apatite 5 from sample CA28 (Appendix S10), in contrast to the other crystals, displays oscillatory thin zoning outside the core. The patterns defined by the oscillatory zoning are parallel to the outer edges of the crystal, clearly outlining edges at different stages of the crystal's growth. The S- and REE-poor cores often have irregular edges suggesting dissolution/re-precipitation events. For example, sample CA28-5 shows a partially resorbed apatite crystal where, oscillatory zones terminate at the resorbed crystal boundary (Appendix S10).

Evidences suggesting a partial replacement of the S- and REE-poor core by dissolution/re-precipitation events are shown in Appendix S10 (*e.g.* samples CA24, CA25, CA28-14). As extreme case, the ovoidal S- and REE-rich region inside the S- and REE-poor core and the outer edges of the S- and REE-poor core seem to have been dissolved away in samples CA24 and CA28-14 respectively (Appendix S10). Such dissolution-related features are often reported in other crystal phases such as plagioclases and olivines (*e.g.* Shore and Fowler, 1996; Milman-Barris *et al.,* 2008).

Transect analysis across crystals (Appendix S11) suggest that apatite crystals from samples CA25 and CA28 are slightly enriched in some of the REE compared with crystals from samples CA10 and CA24. For samples CA25 and CA28 the $Ce_2O_3$ content ranges from 1.0 (core) to 2.0 wt.% (rim), $La_2O_3$ content increases from 0.5 (core) to 1.2 wt.% (rim) and $Nd_2O_3$ content is up to 0.5 wt.% (rim). In samples CA10 and CA24 the $Ce_2O_3$ content varies between 0.6 (core) and 1.3 wt.% (rim), $La_2O_3$ content ranges between 0.3 (core) and 0.7 wt.% (rim), and $Nd_2O_3$ content is ~ 0.3 wt.% (rim). In all the analyzed samples, the $SO_3$ content ranges between 0.7 (core) and 1.2 wt.% (rim).

Overall, transect analyses confirm that chemical zoning is evident in all the analyzed crystals for S, Ce and La whereas Nd zoning is weak in CA25 and CA28 and extremely weak or absent in CA10 and CA24.

*Minor and trace elements by LA-ICP-MS*

*Monovalent cations:* Na concentration is constantly around 1000 ppm and K is around 20 ppm with rims enriched with respect to cores (Appendix S12). Rb and Cs are below the detection limits.

*Divalent cations:* Mg varies between 60 and 500 ppm and it shows core-rim zoning (*i.e.* rims more enriched, Appendix S12). Mn is between 60 and 130 ppm, Fe ranges from 650 to 850 ppm, Ni is less than 1 ppm or below the detection limit as Cu. Ba ranges from 80 to 200 ppm and Pb varies between 25 and 75 ppm. The distribution of these elements is homogenous and no clear core-rim zoning has been detected.

*Trivalent cations:* B ranges from 10 to 70 ppm and it does not show a clear core-rim zoning. Cr and Be are below the detection limit. Y varies between 80 and 600 ppm. Among the LREE, Ce is the most abundant with concentrations ranging from 0.8 to 2.0 wt.% followed by La (0.5 – 1.3 wt.%), Nd (0.2 – 0.6 wt.%), and Pr (0.05 – 0.19 wt.%). The MREE are presents in concentrations varying from less than 10 ppm to 850 ppm: Sm (250 – 850 ppm), Eu (40 – 150 ppm), Gd (120 – 530 ppm), Tb (8 – 50 ppm), and Dy (20 – 175 ppm). The HREE have concentrations ranging from below 1 ppm to 40 ppm: Ho (3 – 23 ppm), Er (4 – 44 ppm), Tm (0.3 – 5 ppm), Yb (2 – 22 ppm), and Lu (0.2 – 2 ppm). Apatite rims are generally slightly enriched in all the REE plus Y respect to their cores (Appendix S12).

*Quadrivalent cations:* Ti concentrations are less than 5 ppm or below the detection limit. Zr is between 7 and 70 ppm and Hf is less than 1 ppm, no zoning related to these elements has been observed. Th ranges from 0.12 to 0.68 wt.% and U is between 140 and 700 ppm. Although Th and U concentrations vary from core to rim no clear zoning has been observed (Appendix S12).

*Pentavalent cations:* Ta is well below 1 ppm whereas Nb is less than 5 ppm. V ranges from 300 to 800 ppm. They are homogenously distributed between core and rim.

*Other elements:* Cl is always below detection limit.

## Discussion

### *Origin of zoning*

Apatite cores would have developed during a first event of isobaric crystallization, where pressure conditions were favorable to the generation of leucite as a major phenocryst and where volatiles were represented by a mixed $H_2O$-$CO_2$ fluid phase as testified by the presence of Ti-garnet and phlogopite (Freda *et al.,* 1997; Gaeta *et al.,* 2000). This crystallization event resulted in the magma becoming oversaturated in volatiles triggering, thus, the Villa Senni eruption (Freda *et al.,* 1997). The removal of magma caused a sudden pressure drop in the magma chamber, which produced a second event of crystallization leading to the development of apatite rims.

### *REE patterns*

The REE patterns show a smooth decrease from La to Lu (Fig. 4). Given that the compatibility of REEs in apatite increase from La to Sm and then decreases toward Lu, this behavior should produce an enrichment from La to Sm (Prowatke & Klemme, 2006).

However, REEs in garnet show a decrease of their incompatibility from La to Sm and become compatible with Sm (van Westrenen *et al.,* 1999). This competitive behavior between apatite and garnet for the REE partitioning permits to develop a continuous smooth decrease in the REE patterns from La to Lu. These evidences suggest that garnet crystallized before apatite.

Apatite from sample CA25 are slightly enriched in REE with respect to apatite from sample CA28 and both have higher contents respect to apatite from sample CA10 that has normalized REE contents similar to those of apatite from sample CA24. Based on the partitioning systematics of the phases on the liquidus of the Colli Albani magma body (*i.e.* clinopyroxene and leucite; Freda *et al.,* 2008) and on the increasing REE distribution coefficient apatite-melt with increasing degree of polymerization of the melt (Prowatke & Klemme, 2006), one would expect increasing REE concentrations on apatite crystallized from more evolved liquids. Consequently, our samples would imply that apatite from sample CA10 and CA24 have crystallized earlier than apatite from sample CA25 and CA28 probably during the initial decompression of the magmatic chamber.

Apatite are characterized by absence of Ce anomalies and presence of negative Eu anomalies. Anomalies in chondrite-normalised REE patterns can be produced by redox transition of $Ce^{3+}$ to $Ce^{4+}$ or $Eu^{3+}$ to $Eu^{2+}$, because the change in valence is associated with the change in ionic radius. Apatite do not show positive Ce anomalies suggesting that extreme oxidation conditions have not characterized their crystallization environment. On the other hand, slightly but evident negative Eu anomalies are found for the apatite studied here with $Eu/Eu^{*}$ values varying in the range 0.52-0.66. Given that $Eu^{3+}$ and $Eu^{2+}$ are both compatible elements in apatite (Prowatke & Klemme, 2006) and considering the mineralogical assemblage of italites composed of garnet + leucite + clinopyroxene + apatite + phlogopite, the only competitor for apatite that could concentrate $Eu^{2+}$ from the melt is phlogopite (Icenhower & London, 1995; Fabbrizio *et al.,* 2010). This evidence suggests that apatite has crystallized after phlogopite.

*Substitution mechanisms*

Apatite from italites of the CAVD is F-bearing (between 2.6 and 3.7 wt.%) and virtually devoid of Cl, characteristic consistent with apatite of magmatic origin (*e.g.* O'Reilly & Griffin, 2000; Patiño Douce *et al.,* 2011; Chen & Simonetti, 2013).

The fluorapatite are characterized by relatively high contents of LREE (1.0 - 4.2 wt.%) and $SiO_2$ (1.3 - 3.6 wt.%). The coupled substitution that follows the scheme $Ca^{2+} + P^{5+} = Si^{4+} + REE^{3+}$ is suggested by the positive correlation (Fig. 5) between LREE contents and Si abundances (Pan & Fleet, 2002; Liferovich & Mitchell, 2006; Chen & Simonetti, 2013). Although the relationship between the Si and LREE abundances falls close to the 1:1 line, the fit of the regression to the data is relatively low (R = 0.93) suggesting that other substitutions can occur.

On one hand, a possibility may be represented by the substitutions involving Na following the schemes $2Ca^{2+} = Na^{+} + REE^{3+}$ and $2S^{6+} + 4REE^{3+} + Si^{4+} + 2Na^{+} = 4P^{5+} + 5Ca^{2+}$ (Liu & Comodi, 1993; Parat *et al.,* 2002). However, the roughly constant $Na_2O$ contents at ~ 0.1 wt.% shown by the analyzed apatite seem to rule out these substitution mechanisms. The high $SO_3$ contents of apatite (Appendix S2) could indicate the presence of the substitution mechanism $S^{6+} + Si^{4+} = 2P^{5+}$. However, the absence of a 1:1 correlation between $S^{6+} + Si^{4+}$ and $2P^{5+}$ abundances indicates that additional substitution mechanisms should have played an important role (Fig. 5). We note that the relative high Th concentrations (up to 6800 ppm, Appendix S2) suggest that $Th^{4+}$ may be involved in the substitution processes following the scheme $Th^{4+} + 2Si^{4+} = Ca^{2+} + 2P^{5+}$ (Chen & Simonetti, 2013). A plot of the analytical data in a ($Th^{4+} + 2Si^{4+}$) versus ($Ca^{2+} + 2P^{5+}$) diagram (Fig. 5) clearly shows a very good correlation ($R = 0.98$), suggesting that the substitution mechanism involving $Th^{4+}$ may play an important role for the apatite from italites of CAVD.

*Comparison with apatite from Italian alkaline rocks*

A comparison from other Italian alkaline rocks evidences as apatite compositions can vary widely, from apatite poor in Sr, REE and halogen to apatite extremely enriched in REE, Th and U.

*Apatite from Colli Albani rocks:* Apatite found in xenoliths ejected during the latest activity of CAVD are characterized by significant concentrations of Si, S, Sr, La, Ce, Th and F ($SiO_2$ = 1.56-4.46 wt.%; $SO_3$ = 0.81-3.74 wt.%; SrO = 0.70-4.04 wt.%; $La_2O_3$ = 0.55-2.75 wt.%; $Ce_2O_3$ = 0.87-4.26 wt.%; $ThO_2$ = 0.44-6.65 wt.%; F = 3.86 wt.%; Federico & Peccerillo, 2002). Apatite found in the groundmass of leucitite lava flows of CAVD (Melluso *et al.,* 2010) has lower $SiO_2$ (1-2 wt.%), SrO (0.8-1.0 wt.%) and LREE (0.45-2.0 wt.%) contents than those studied here, whereas the F contents are in the range 4.3-5.3 wt.%, well above the maximal F content of 3.77 wt.% for stoichiometric Ca-fluorapatite, suggesting a potential overestimation due to the increase in apatite F X-ray intensity during EMP analyses related to the use of the energy-dispersive spectrometry system (Pyle *et al.,* 2002).

*Apatite from ultramafic rocks:* Apatite from different ultra-alkaline rocks from central and southern Italy, namely from melilitite of Cupaello and from lavas of Vulture (tephrite and basanite-foidite), have lower $La_2O_3$ (0.03-0.38 wt.%), $Ce_2O_3$ (0.20-0.67 wt.%), and SrO (0.36-0.67 wt.%) contents, $Na_2O$ up to 0.19 wt.%, $SiO_2$ in the range 0.9-1.35 wt.%, $SO_3$ up to 1.3 wt.%. Apatite from Cupaello have F around 3 wt.% and low Cl contents (< 0.1 wt.%), whereas those from Vulture have F contents in the range 1.2-1.7 wt.% and Cl up to 0.55 wt.% (Stoppa & Liu, 1995). Apatite crystals found in a kalsilite-bearing leucitite from Abruzzi, Italy, show core-rim zoning with strong enrichment in $SiO_2$ (4-5 wt.%), SrO (2 wt.%), $SO_3$ (1.9-2.6 wt.%) and LREE (4-7 wt.%) in the rims compared to cores, $SiO_2$: 1-2 wt.% - SrO: 0.6-1 wt.% - $SO_3$: 1.6-1.9 wt.% LREE: 0.6-1.1 wt.%, (Comodi *et al.,* 1999). Apatite found in the leucite bearing intrusive rocks (i.e. fergusites) from the crystallizing sidewalls of Vesuvius magma chambers have low SrO content (0.20-0.45 wt.%), high F (3.25-3.51 wt.%) and Cl (0.43-0.71 wt.%) concentrations (Fulignati *et al.,* 2000). Apatite crystals from the groundmass of the S. Caterina melafoidite lava, Mt. Vulture, although their F contents are overestimated (F = 3.8-5.0 wt.%), are characterized by relative high concentrations of $SO_3$ (up to 2 wt.%) and Cl (0.2-0.3 wt.%), low SrO (0.3-0.7 wt.%) and LREE-oxides contents (0.2-1.5 wt.%) and are commonly accompanied by a britholite rim (Melluso *et al.,* 2011).

*Apatite from intermediate rocks:* Apatite found in phonolite from central Italy in the localities of Titignano and Macchie have low $La_2O_3$ (up to 0.16 wt.%), $Ce_2O_3$ (0.14-0.27 wt.%), and SrO (0.33-0.56 wt.%) contents, $Na_2O$ up to 0.12 wt.%, $SiO_2$ in the range 0.9-1.3 wt.%, $SO_3$ up to 1.05 wt.%, F around 2.7 wt.% and Cl up to 0.13 wt.% (Stoppa & Liu, 1995).

The most interesting finding of this comparison is that fluorapatite from italites of CAVD are the Sr-richest apatite in the context of the Italian alkaline rocks. As divalent cation Sr can enter the apatite structure substituting for Ca (Pan & Fleet, 2002). This enrichment in Sr contents is clearly related to the mineralogical assemblage of italites that does not have Sr-compatible mineral phases other than apatite. In all the others Italian alkaline rocks, with the exception of the kalsilite-bearing leucitite from Grotta del Cervo (Comodi *et al.,* 1999), apatite is accompanied with mineral phases, such as sanidine, plagioclase, and melilite, where strontium is a compatible element (Stoppa & Liu, 1995; Fulignati *et al.,* 2000; Melluso *et al.,* 2010; Melluso *et al.,* 2011). The crystallization of these Sr-compatible mineral phases reduces the amount of Sr available in the residual liquid for apatite, producing thus apatite with low SrO contents (≤1 wt.%). This hypothesis is also supported by the fact that apatite found in the kalsilite-bearing leucitite from Abruzzi is relatively enriched in SrO (up to 2 wt.%), and by the lack of Sr-compatible mineral phases in the hosting rock (Comodi *et al.,* 1999). The higher SrO contents, up to 4.6 wt.%, in apatite from italites of the CAVD respect to those of apatite from the kalsilite-bearing leucitite of Grotta del Cervo can be explained by Sr assimilation from the surrounding sedimentary carbonate wall rock as suggested by oxygen isotope data in mineral phases from various rock-samples (Federico *et al.,* 1994; Federico & Peccerillo, 2002; Dallai *et al.,* 2004).

## *Conclusions*

The high F contents and virtual absence of Cl confirm the textural evidences of a primary magmatic origin for the studied apatite and exclude the possibility that they could have crystallized in a hydrothermal environment. Despite the chemical composition of apatite from other Italian alkaline systems can vary widely, the F-apatite in the italites of the CAVD are the richest in SrO contents. This peculiarity is strictly related to the absence of Sr-compatible mineral phases, other than apatite, such as plagioclase, melilite, and sanidine in the mineralogical assemblage of italites from the CAVD and to Sr assimilation from the surrounding sedimentary carbonate wall rock. Moreover, the presence of core-rim zoning with rims clearly enriched in LREE and $SO_3$ suggests that it was developed during a massive event of crystallization induced by the pressure drop in the magmatic chamber during the eruption. In addition, the absence of a positive Ce anomaly suggests that the S enrichment cannot be due to variable oxidation conditions during the crystallization of the magma body. The structure of apatite hosts S and REE substituting for Ca and P and provoking thus a reverse core-rim zoning for Ca and P. As a consequence of these substitution mechanisms Si is forced inside apatite to balance the electrovalence provoking the Si enrichment at the rims of the crystals. Finally, the different levels of enrichment in REE contents suggest that apatite crystallized at different stages during the eruption.

## **Acknowledgements**

We thank Gianluca Iezzi and an anonymous reviewer for their constructive comments. This manuscript is an extended version of the unpublished Tesi di Laurea in Scienze Geologiche (N.O.) of the first author. He wants to thank his parents for having given him the possibility to study at the highest level. Martina Tribus is thanked for her assistance with the electron microprobe. MP acknowledges the support of the ERC grant 612776 CHRONOS (Diego Perugini principal investigator).

**Figure captions**

Fig. 1. Location and geological sketch map of the Colli Albani Volcanic District. Three main periods of activity are: (I) Tuscolano-Artemisio (or Vulcano Laziale lithosome, Giordano et al. 2006), dominated by large explosive eruptions; (II) Faete, mainly characterised by strombolian and effusive activities, and (III) Hydromagmatic, including monogenetic and polygenetic maar-forming eruptions. Sample locality (Fosso di fioranello) of italites is indicated.

Fig. 2. Field aspect of the italites and upper flow deposit (*Pozzolanelle)* of the Villa Senni eruptive unit at Fioranello locality. (A): detail of the typical "pozzolane" texture, *i.e.* a poorly lithified, massive, coarse-grained deposit of scoria lapilli and blocks (the red tape measure is about 20 cm in diameter); (B): block-sized italite (sample CA10), *i.e.* a granular, leucocratic, lithic clast. (C): detail of the italite texture (sample CA10); most of the black and white minerals are clinopyroxene and leucite respectively.

Fig. 3. Photomicrographs of italite thin section (sample CA10) in crossed polars light. The photos show the holocrystalline, granular texture of italite rocks made up of subhedral clinopyroxene (Cpx), anhedral leucite (Lct), subhedral phlogopite (Mca), anhedral garnet (Grt) and subhedral apatite (Ap; note the millimetre sized apatite crystal in the centre of photo B).

Fig. 4. Chondrite-normalised REE abundances of representative apatite, cores (solid lines) and rims (dashed lines). Chondrite values are from McDonough and Sun (1995).

Fig. 5. Plots illustrating a positive, linear relationship between: (a) Si and LREE atomic contents (all apatite); (b) S + Si and 2P atomic contents (all apatite); (c) Ca + 2P and Th + 2Si atomic contents (selected apatite; *i.e.* apatite analyzed by LA-ICP-MS).

Figure 1

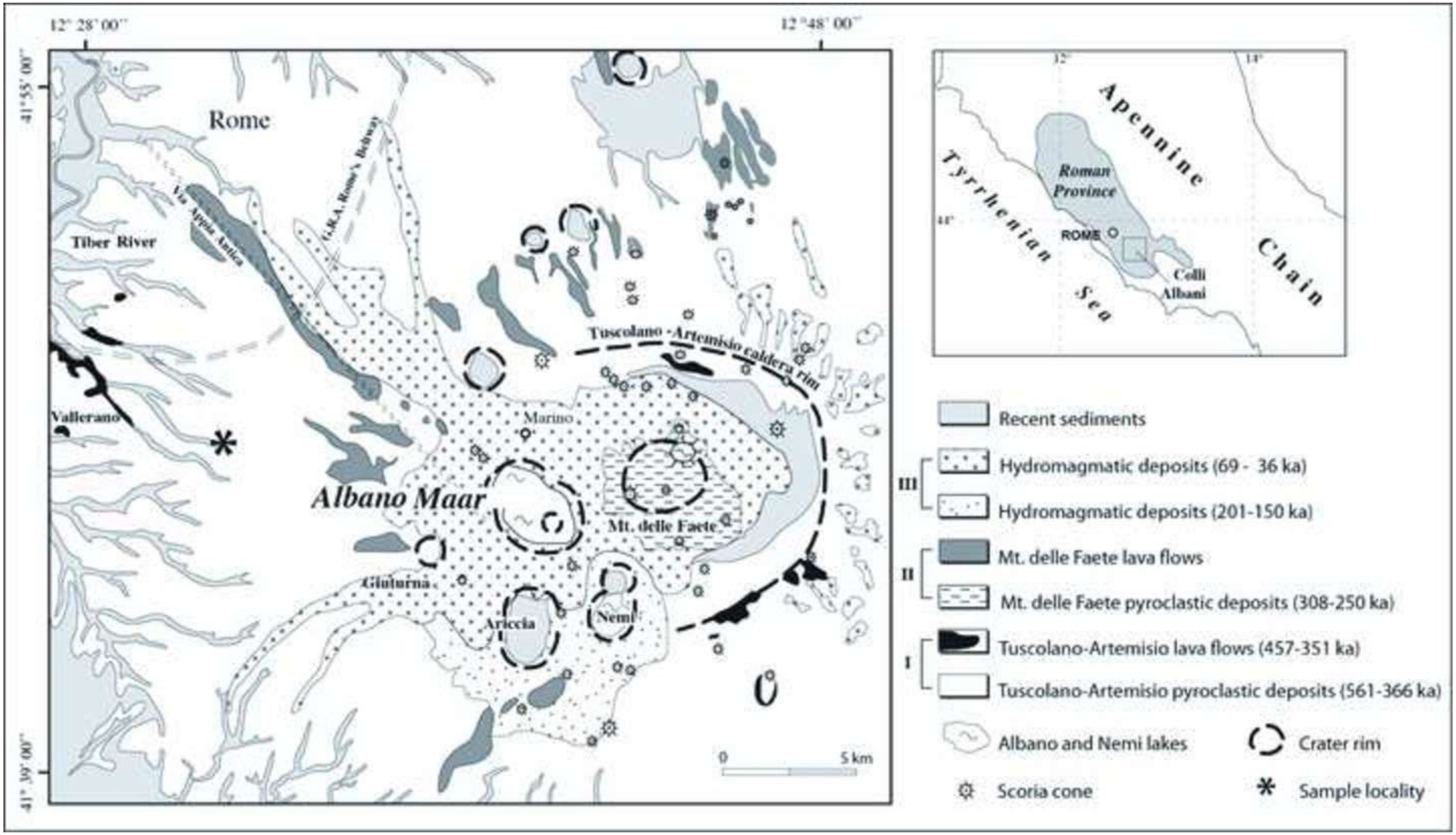

Figure 2

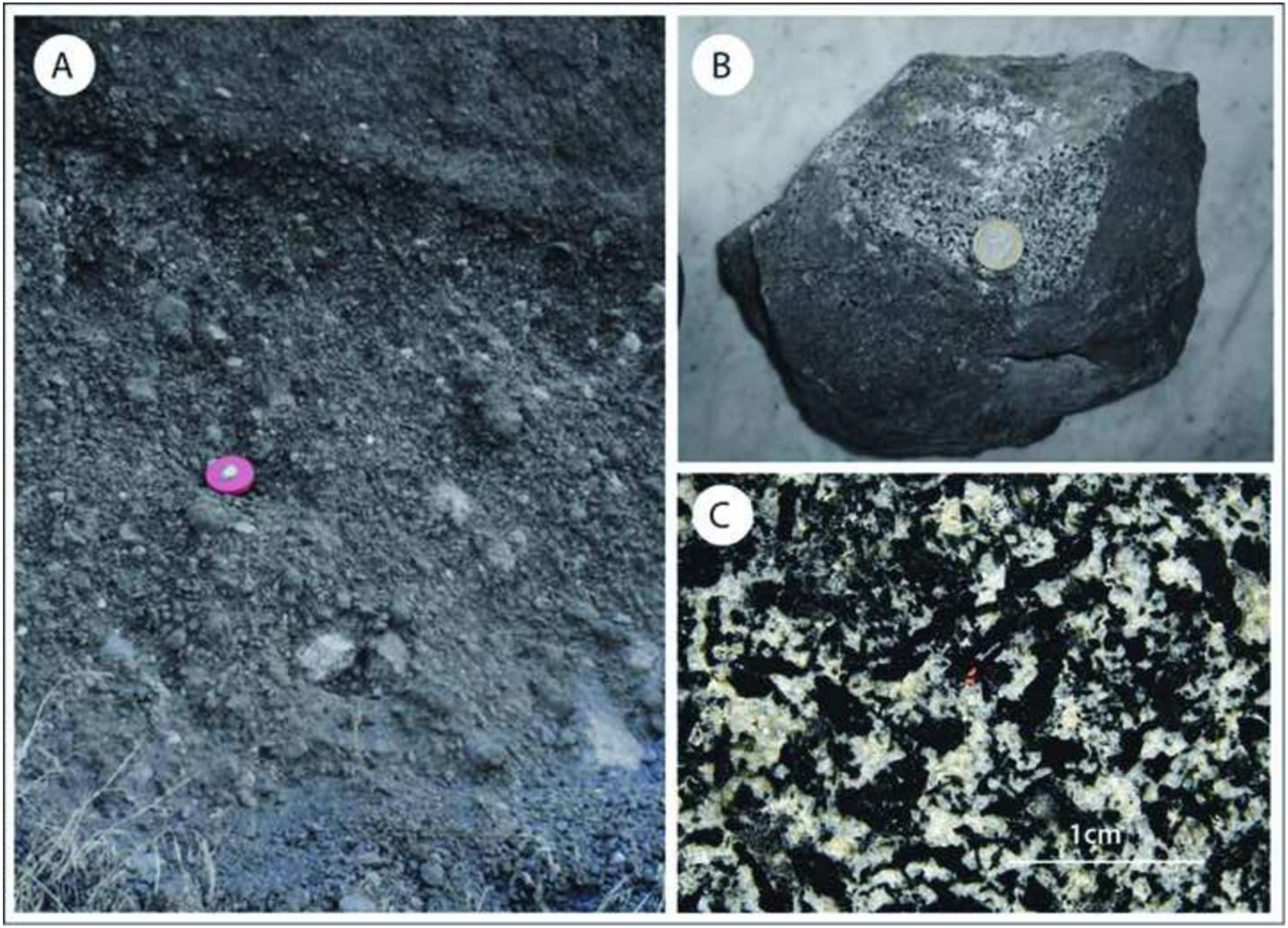

Figure 3

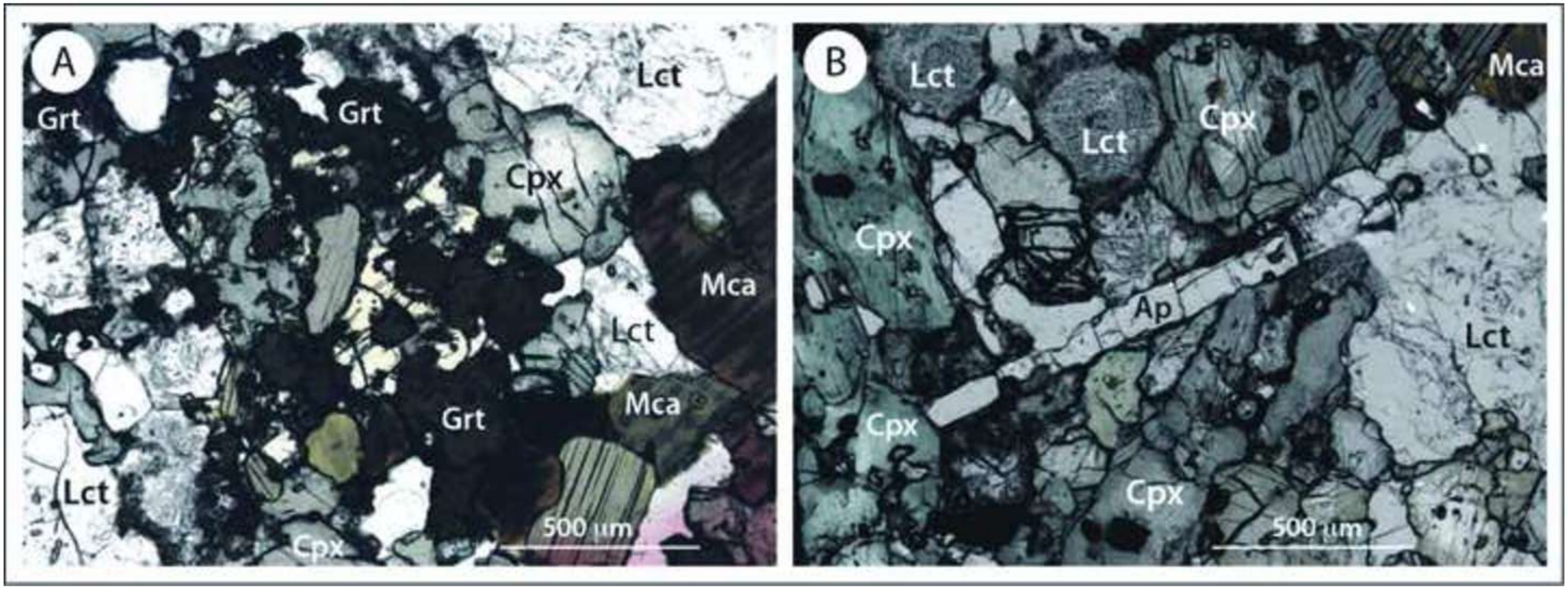

Figure 4

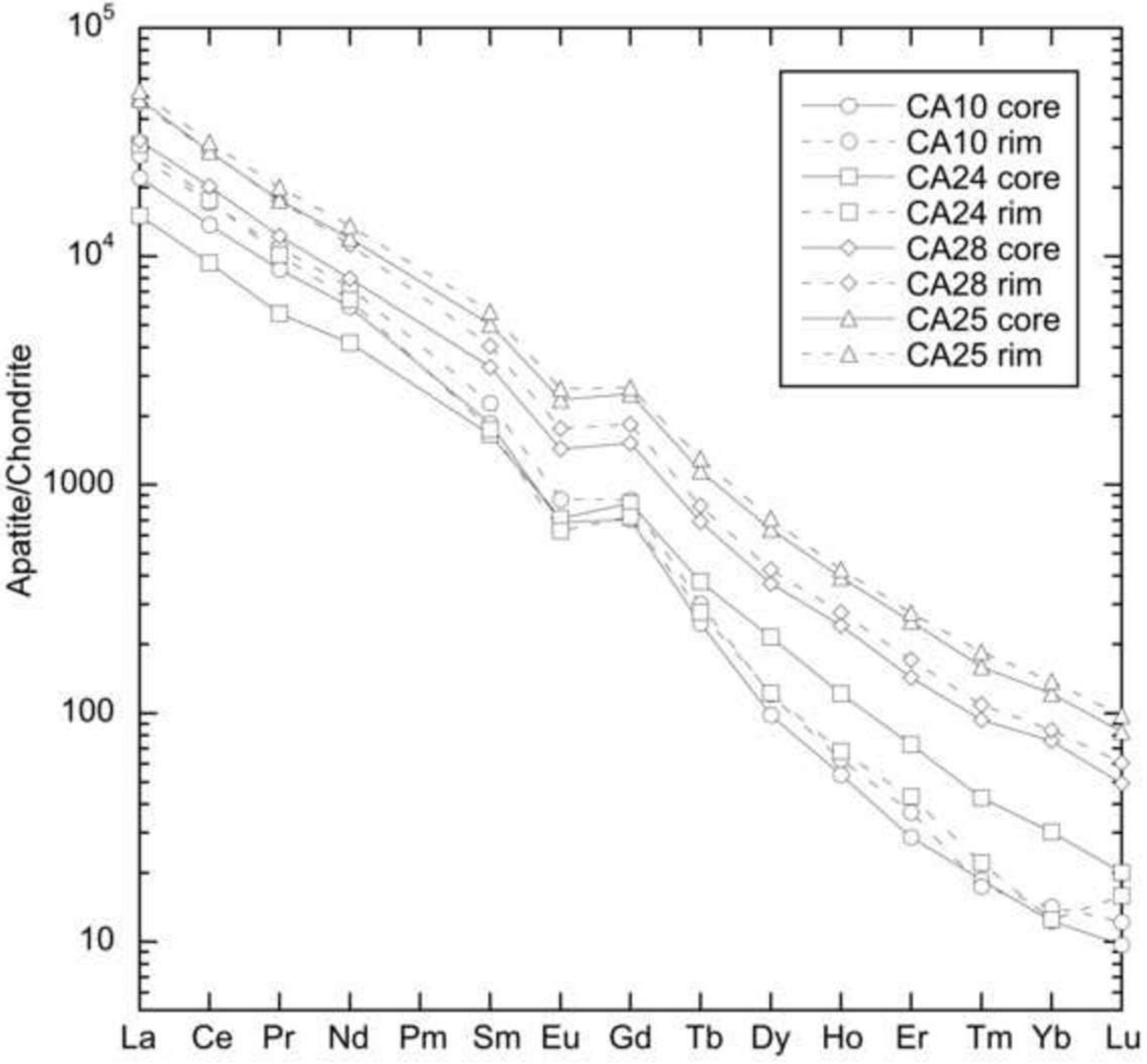

Figure 5

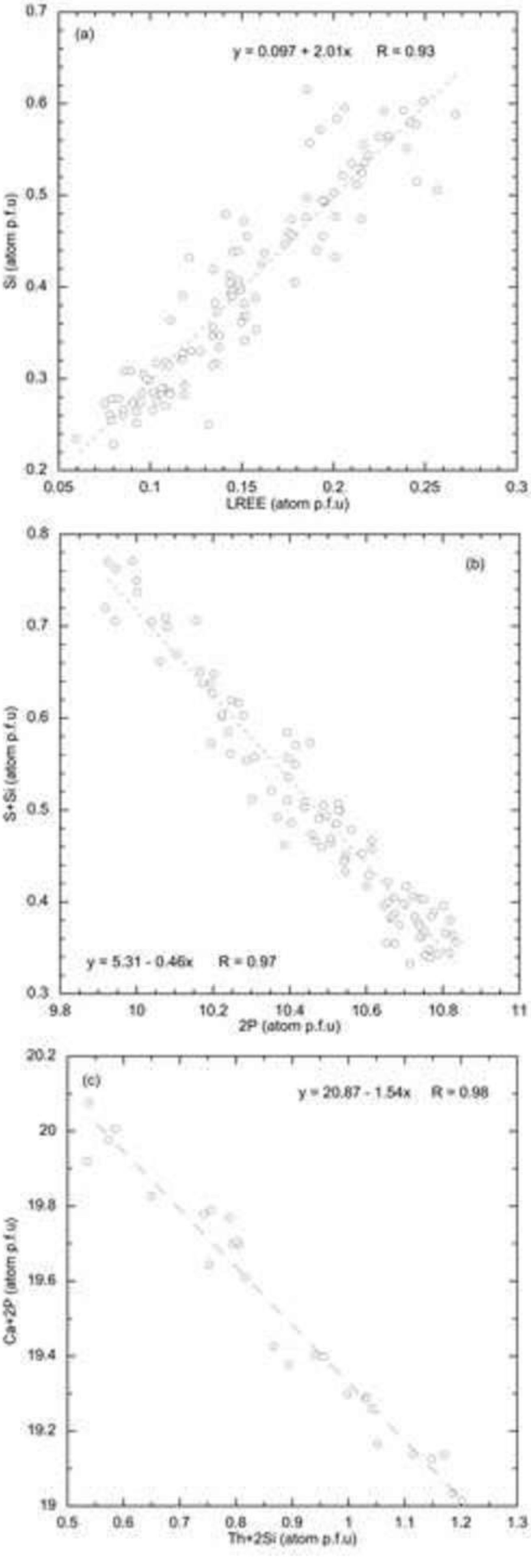
(a)
y = 0.097 + 2.01x R = 0.93
Si (atom p.f.u)
LREE (atom p.f.u)
0.2
0.3
0.4
0.5
0.6
0.7
0.05
0.1
0.15
0.2
0.25
0.3
(b)
y = 5.31 - 0.46x R = 0.97
S+Si (atom p.f.u)
2P (atom p.f.u)
0.3
0.4
0.5
0.6
0.7
0.8
9.8
10
10.2
10.4
10.6
10.8
11
(c)
y = 20.87 - 1.54x R = 0.98
Ca+2P (atom p.f.u)
Th+2Si (atom p.f.u)
19
19.2
19.4
19.6
19.8
20
20.2
0.5
0.6
0.7
0.8
0.9
1
1.1
1.2
1.3